# Evaluation of a New Vasculature by High Resolution Light Microscopy: Primo Vessel and Node

Vitaly Vodyanoy,[1,2*] Oleg Pustovyy,[1] Ludmila Globa,[1] and Iryna Sorokulova[1,2]

[1]Department Anatomy, Physiology and Pharmacology, College of Veterinary Medicine Auburn, AL 36849; [2]School of Kinesiology, Auburn University, Auburn, AL 36849.

\* vodyavi@auburn.edu

## Abstract

In the 1960's Bong Han Kim discovered and characterized a new vascular system. He was able to differentiate it clearly from vascular blood and lymph systems. He demonstrated that this system is composed of nodes and vessels, and it was responsible for tissue regeneration. These findings remained dormant for many years until recently when a new interest in this anatomical vascular system appears. The system is now termed primo vascular system. In this work, we characterized microstructure of rodent primo nodes and vessels using a high resolution light microscopy. The most typical primo nodes are oval-shape of 0.1-0.5 mm over the short and 0.5-1 mm over the long axis. Both node's ends are linked to primo vessels of 3-6 cm long and 40-100 µm in diameter. The primo vessel is composed of 1–20 subvessels of 3–25 µm in diameter. The external envelope of the subvessel is composed of two layers: the wall of endothelial cells with a rod-shaped nucleus of 15-20 µm and the outer membrane containing spindle-shaped cells with ellipsoidal nucleus of 13-27 µm long and 4-5 µm thick, that are similar to smooth muscle cells. The subvessels are surrounded by fine, longitudinal and circular fibers crossing each other. The bundle of subvessel of the primo vessel is laid into an external jacket composed of endothelial cells with 6–12 µm round or oval nuclei. The node is heterogeneous in nature, composed of twisted subvessel bundles that fill up nearly the entire node volume. The enlarged subvessel inside the node harbors microcells that express stem cells and stem cell niche markers. We conclude that these microcells are progenitors of multipotent stem cells and the nodes serve as the stem cell niches outside the bone marrow.

## Introduction

The idea of the conduit harboring hematopoietic stem cell independent of the bone morrow was proposed by Alexander A. Maximow, the scientist who coined the name "stem cell" [1]. He described structures and functions of hematopoietic systems as scaffolds or niches for stem cell maturation and subsequent development of blood cells: "*Being endowed with ample prospective potencies they can produce, provided external conditions are favorable for hemocytoblasts* (hematopoietic stem cells) *and different types of blood cells… remaining throughout the whole life in an undifferentiated embryonic condition. …They represent a vast cell system, distributed all over the body, over various organs and assuming, according to their position, manifold histological aspects.*"[2].

The existence of a new circulatory system that is different from blood and lymphatic vasculature was reported by Bong Han Kim, a North Korean scientist and professor of Pyongyang Medical College in the 1960's. He stated that this system, he named Bonghan system, is composed of nodes and vessels, and it was responsible for tissue regeneration [3-8]. Kim described the isolated microcells from the newly discovered vascular system and induced their proliferation under artificial conditions. He stated that microcells had a cell-like structure, contained chromosomes, and participated in the tissue regeneration [9]. Each of these properties belongs to stem cells [10]. Bong-Han Kim also demonstrated that microcells were harbored in a special node, a highly vascular organ that provides physiological conditions favorable for microcells [9], as well as a stem cell niche [11]. The work of Bong Han Kim was recently reviewed [12, 13].

According to Kim, primo vascular system is composed of nodes and vessels. Each primo vessel connects primo nodes together, and each primo node is linked with primo vessels (Figure 1). The primo vessel comprises of a bundle of subvessels. The subvessel bundle of the incoming (afferent) vessel enter the node, branches into additional bundles, and fills the node interior by tightly spun and folded bundles. The subvessels narrow, converge, and come out from the node as a single bundle of the efferent primo vessel. The vessels are covered with the external jacket that gradually turns into the node capsule [9, 13, 14].

In 2002, the scientific group of Dr. Kwang-Sup Soh initiated a series of experiments that validated many of Kim's results. This research has ignited a new interest in this anatomical vascular system that now termed primo vascular system (PVS) [12]. Critically, the primo vascular system was identified as a potential stem cell niche for multipotent stem cells outside of the bone morrow [15-18].

The structures of primo node and vessels were characterized by a variety of physical and biochemical methods, including electron [19-25], atomic force [23], confocal [26-29], fluorescent [21, 30, 31], and light [32-35] microscopies. It was demonstrated that these structures are different from the lymphatic and blood vasculatures [36, 37]. However, internal structure of primo node and detail structure of the intra-node primo vasculature were not fully characterized. In our previous work, the structure of primo nodes were initially described [38]. In this work, we use high resolution light microscopy to evaluate primo nodes and subvessel sinuses as potential sites that harbor stem cell progenitors.

## Material and Methods
### Microscopy
It was recently demonstrated that 90 nm resolution in the images was achieved by using an optical illumination system with a high-aperture cardioid annular condenser [39]. The high resolution of the optical system is accomplished by two special features of the system. First, the effect of the annular illumination is the narrowing of the central spot of the diffraction pattern and increase of the intensity of the diffraction fringes. A similar effect of annular aperture is known and used in telescopes for the resolving power increase [40]. Second, the illumination system produces a coherent illumination of small (compared to wavelength) object and thus the coherent image. Hence, partial cancellation of the diffraction amplitude occurs when the antiphase subsequent fringes of diffraction are added in the image. This property makes the image edge sharper and thus increases the resolution [39]. The structural annual illumination gives an additional advantage to this system in that provides the capability to produce optical sectioning of thin samples similar to that found in confocal microscopy [41]. The optical sectioning allows for the discerning of in-focus image from out-of-focus structures. It allows the user to see not only the sample cutting surface, but also makes it possible to glance inside optical sections below the cutting plane. Fine focusing and placement of the focus in any depth of the sample allows for positioning of the focus at a desirable increment. In this manner, a three-dimensional profile of the sample can be obtained. The system was successfully tested for observation of still and motile cells [42], nanoforms and vesicles produced by fragmented erythrocytes [43-45], and by characterization of primo nodes and vessels [14].

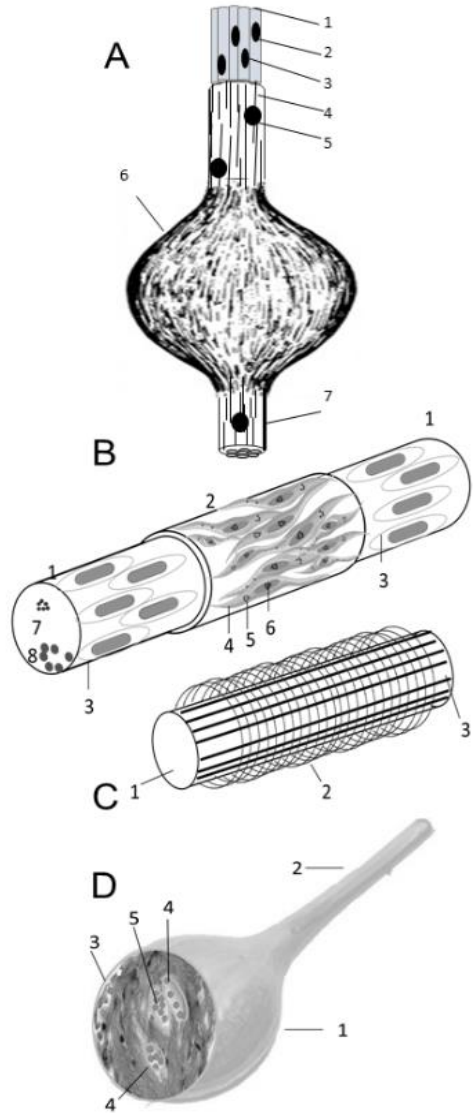

Figure 1. Primo node, vessel, and subvessels. A. Primo node, incoming and out coming vessels; 1-primo-subvessel; 2-cell nucleus of the outer membrane; 3-nucleous of endothelial cell; 4-external jacket of the out coming primo-vessel; 5-nucleous of jacket endothelial cell; 6-primo node; 7-out coming primo vessel [7]. B. Diagram of primo-subvessel. 1-wall of subvessel formed by endothelial cells; 2-outher membrane of subvessel; 3-endothelial cell with rod-shaped nucleus; 4-spindle-shaped cell with ellipsoidal nucleus; 5-fine basophil granules in the cytoplasm; 6-fine chromatin granules inside nucleus; 7-basophil granules inside the subvessel; 8-p-microcells. C. Diagram of subvessel fibers. 1-primo subvessel; 2- fine transversal fiber; 3-longitudinal fiber. D. Diagram of the transversal section of a primo-node. 1. Primo-node; 2 – primo-vessel; 3 – node capsule; 4 – lumens, 5 – p-microcells.

## Sample preparation

Materials and sample preparation were explained in detail in previous work [84]. In short, the primo node and vessel samples were collected from the surfaces of the intestine in the peritoneal cavity of the Sprague–Dawley rats. Samples collected from rats were analyzed in 2 different ways. For images of unfixed samples with high resolution light the microscope, the primo nodes and vessels were removed and submerged in a phosphate buffer solution (PBS, pH = 7.4) stained with 0.01% acridine orange. Other tissues were fixed in Bouin's fluid, embedded in paraffin, sectioned at 6 mm slide, mounted, H&E (hematoxylin and eosin)-stained, and covered with a cover slide.

## Ethics Statement

Approval was obtained from the Auburn University Institutional Animal Care and Use Committee for performing this study.

## Results and Discussion

### Primo nodes and vessels

High resolution light microscopy allows for the production of highly informative images of primo vascular nodes and vessels. Figure 2 shows

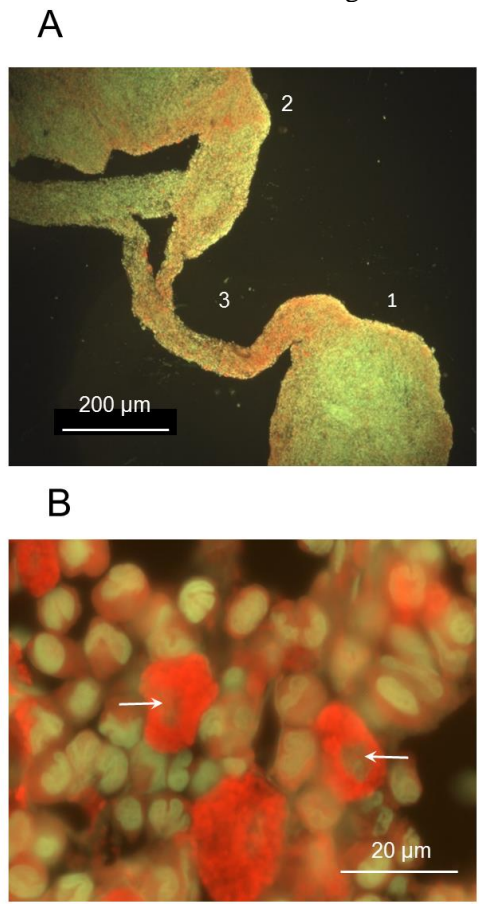

Figure 2 Non-fixed intra-external primo nodes and primo vessels from a rat, stained with acridine orange: A. A primo node 1 is connected with a primo node 2 through the branching primo vessel 3. B. A magnified view of small surface of the primo vessel. Mast cell nuclei are shown by arrows

primo-vascular nodes and vessels harvested from the surface of the intestine. The acridine orange staining reveals cells fluoresce with green and red light. Figure 2 illustrates this case, where one end of a node (Figure 2 A (1)) is connected to the primo-vessel (3). This primo-vessel is branched and connected to another node (Figure 2 A (2)) [9]. The general architecture of primo node connected to primo vessel agrees with a general diagram of the system shown in Figures 1 A and D.

The initial impression of the colors in Figure 2 is that the green and red colors belong to DNA and RNA, respectively, of cell nuclei stained by acridine orange [46]. However, acridine orange also accumulates and emits red light in mast secretory granules and other cellular acidic compartments [47, 48]. Figure 2 B shows that the red color belongs to cells rather than to cell nuclei. The nuclei of the stained red cells are green (arrows show nuclei). The red stained cells represent ~ 5% of cell population. The most likely explanation is that the red stained cells are mast cells. This suggestion is consistent with the fact that elevated amount of mast cells are found in both under acupuncture meridian lines and in primo nodes and vessels [35, 49-52]. The microscopic observation of the mast cells located under the acupuncture meridian lines of humans and rats revealed that the mast cells were more concentrated under the meridian lines in comparison with their control areas [49]. Similarly, the density of mast cells collected from a rat skin at the ST36 acupuncture point was found to be higher than that from a nearby control point. In addition, stimulation of the point resulted in a significant increase in degranulation of the mast cells [50]. Masts cells were found in primo-nodes harvested from the inside of lymphatic vessels and the surface of the intestine. The mast cell amounted for up to 20% of the other primo-node cells. The substantial number of mast cell granules were found in primo subvessels [51]. Mast cells were found in primo-nodes obtained above the epicardia of rat hearts. Some of the mast cells were observed to degranulate into the pericardial space [35].

Our results shown in Figure 2 B agree well with those obtained with primo-nodes and vessels harvested from rat internal organ surfaces. When these samples were analyzed with light microscopy and acridine orange, mast cells appear as red cells with green nuclei [35]. The acridine orange staining of mice peritoneal mast cells also revealed red dyed cells indicative of the presence of acidic granules [48].

These results are consistent with Kim's discovery of "granulopoietic "cell line in PVS that include a granulocyte, mast cell [8, 9].

## Comparison of vessels in primo node, lymphatic node, and blood vessels

In comparing p-subvessel (primo vascular subvessel) with other vascular structures, Kim stated: "*The characteristics of the ductules are distinguished clearly from those of the nerve fibers, connective tissue fibers, blood and lymphatic vessels*" [9]. Our observations of vessels inside the primo and lymphatic nodes and blood vessels are consistent with this statement.

Figure 3 shows transversal sections of a small primo node, lymphatic node, and blood vessels of the rat. Figure 3 A depicts the darkfield image of the section of the small primo node of 220 ×400 nm. A node consists of a bundle of p-subvessels. The section shows p-subvessel lumens (L) of different sizes. The image shows fiber structures that surround subvessels and go spirally around the perimeter of the node. The walls (w) of subvessels are smooth and uninterrupted, except for the wall of the lumens labeled by white arrow. There are small openings in the wall that lead to vacuoles (white arrow). Some p-subvessels run very close to each other with a small intravascular space (white arrow head). The section also shows randomly scattered single cells and granules (black arrow). The node is covered by the outside capsule (C). The structure of the primo node slice is consistent with the general diagram of primo node section shown Figure 1 D. The features of the primo-node cross section image shown in Figure 3 A generally agree with those described by electron and light microscopies [36, 53].

The cross section of the mesenteric lymph node of a rat is shown in Figure 3 B. Lumens of the lymph node are more irregular and wider when compared to the lumens of primo nodes. The primo node lumen is very different from the cross section of blood vessels obtained from a small intestine (Figure 3 C). The comparison of the distinguishing features of primo vessels, blood and lymphatic capillaries were characterized by electron microscopy [36]. The walls of the p-subvessels were found to contain endothelial cells with distinctive rod-shaped nuclei and were not enclosed in a basal lamina or by accessory cells, such as pericytes or smooth muscle cells. It was also found that p-subvessels were not fastened by anchoring filaments to the fibers of extracellular matrices as seen in lymphatic capillaries. The authors concluded that p-subvessels, blood and lymphatic capillaries are significantly different [36].

### Internal structure of primo node

Kim described the primo node as the anastomosis of widened and branched p-subvessels. A p-vessel (primo vascular vessel) bundles fill the node interior by snugly spun and folded bundles [9]. Figure 4 shows longitudinal sections of primo node. The primo node was obtained from the surface of a small intestine. It was of oval shape of 1,600×760 µm. The node was cut into 127 longitudinal slides of 6 µm thick.

The node is heterogeneous in structure, comprised of twisted p-subvessel bundles that fill almost the whole node volume. Vascular bundles interlace in such close-fitting structure that node maintains its physical integrity even without the external capsule that was lost during sample preparation (Figure 4 A). The image of the node is consistent with that shown in [14].

The magnified part of the primo node (Figure 4 B) likewise demonstrates that the internal pattern of p-subvessels sustains the bundle-like design. The bundle of primo-capillaries ($B_∥$) on the right side of the section runs along whole node body. The diameter of p-vessels in this bundle is ~3 µm. The bundles ($B_∥$) economically pack the node volume by serpentine-like structures. The top part of this sample shows the transversal cross sections of vascular bundles ($B_⊥$), which are comparable in appearance to patches of straw stubble. These patches resulted from cutting vessel strands that are vertical with respect to the section plane. The cross section of two vessel sinuses surrounded by fibrous tissue is an important feature of this section (dotted rectangle).

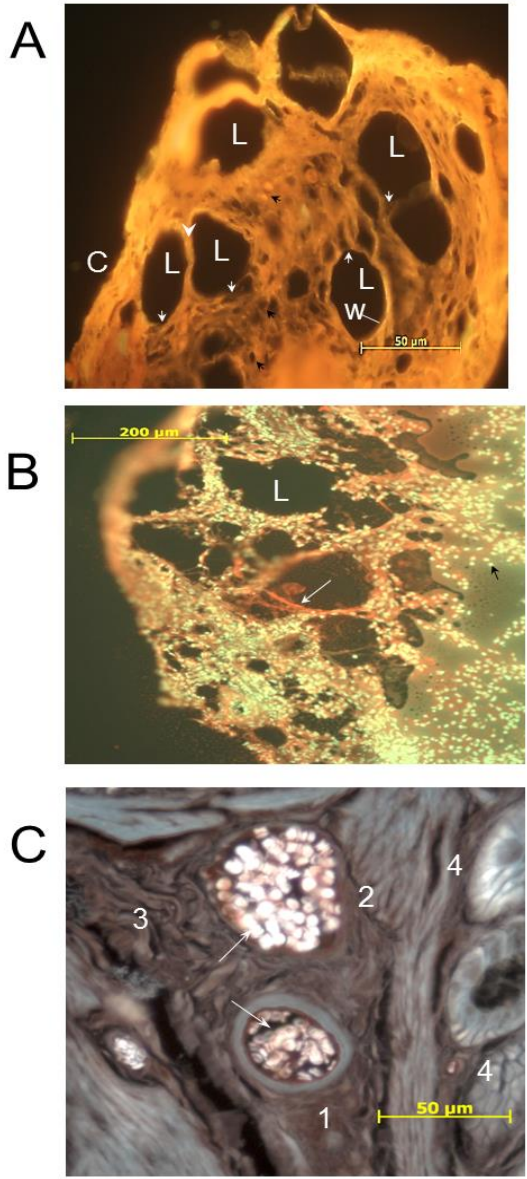

Figure 3. Transversal section of a small primo node, lymphatic node, and blood vessels of rat. A. Primo node harvested from the surface of a small intestine. L - lumens of p-subvessels inside the node; w – wall of p-subvessel surrounded by fibers; C – node capsule; granules -black arrows; the white arrow shows the interrupted portion of lumen connected to a vacuole; scale bar 50 µm. B. Cross section of mesenteric lymph node. L – lumen, white arrow – fiber, black arrow – lymphocyte; scale bar 200 µm. C. Cross section of blood vessels obtained from a small intestine. 1 – artery, 2 – vein, 3 – fiber, 4 – intestinal crypts, white arrows – erythrocytes; scale bar – 50 µm

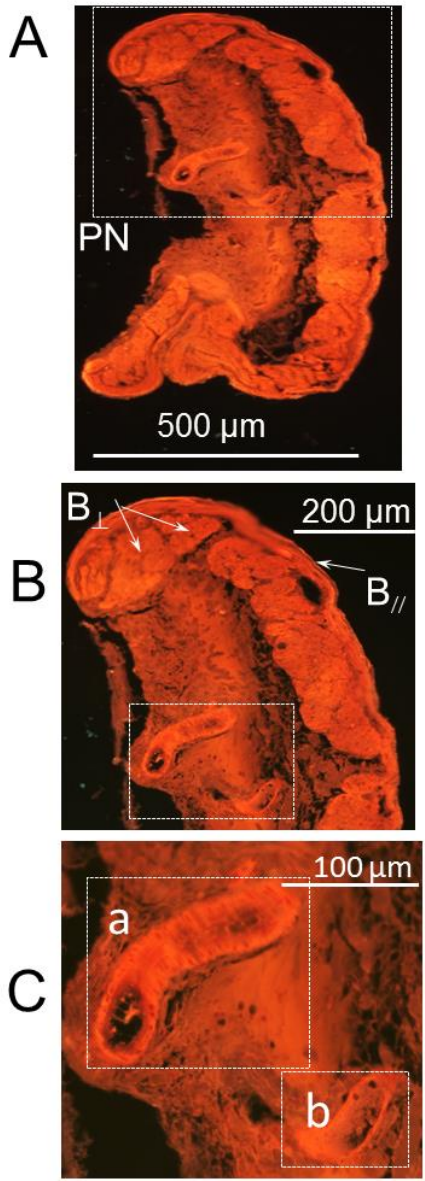

Figure 4. Longitudinal sections of primo node from the surface of a small intestine. A. The section represents the 66th slide of this series showing that primo node (PN) is a complex anastomosis of vessels. B. Magnified view of the dotted rectangle in Figure 4 A. The internal laying of vessels conserves the bundle-like structure. B∥ is the longitudinal bundle of primo-capillaries; B⊥ is the cross sections of transversal vascular bundles. Two vessel sinuses surrounded by fibrous tissue are shown in the lower portion of the slide (dotted rectangle). C. Magnified view of the rectangle in Figure 4 B. Rectangle a – top broaden p-subvessel (sinus), b – lower p-subvessel sinus.

Figure 4 C depicts the magnified area shown in the dotted rectangle of Figure 4 B. The cutting plane of this slide passes through the lower part of the first sinus (rectangle *a*) while the second sinus (rectangle b) is cut near the top of the vessel. In conjunction with these facts, the optical sectioning of the first sinus allows to visualize cellular components of the external envelope of p-subvessel and fibers. On the other hand, the optical sectioning of the second sinus affords the ability to view inside of the sinus.

Originally, the structures of primo-nodes were carefully described by Bonghan Kim. He recognized a few different types of fibers. The primo subvessels in the outer layer are surrounded spirally by argyrophilic fibers, and elastic fibers in this layer run longitudinally. The main organization of the inner structure is tubular one formed by enlarged primo subvessels which are called the sinuses of the node. The p-node is covered with the outer membrane while inside the node, fibrous connective tissue is found between sinuses. This connective tissue contains collagenous, elastic and argyrophilic fibers [9]. The tubular structure and fibrous nature of the rabbit intra-extravascular primo nodes were confirmed by light and transmission electron microscopy (TEM) [19]. The argyrophilic character of some of the fibers was later shown by Lee et al. [25]. Use of magnetic nanoparticles, confocal microscopy, TEM, cryo-SEM and focused-ion-beam SEM, visualized the reticular fibers in the extracellular matrix of primo node inside the rat lymphatic vessel and internal organ surfaces. Images revealed collagen fibers, which formed the extracellular matrix of nodes and vessels. Sinuses of various diameters, which were cross-sections of the lumens of the subvessels, an immune-function cells, like macrophages, mast cells, and eosinophils were also observed [21, 24, 25, 54-56].

## Primo node sinus – enlarged p-subvessel

Primo subvessels are normally quite thin and range of 3-25 µm in diameter [9]. Primo vascular sinuses are large than regular p-subvessel and therefore the sinuses are much accessible for imaging by light microscopy.

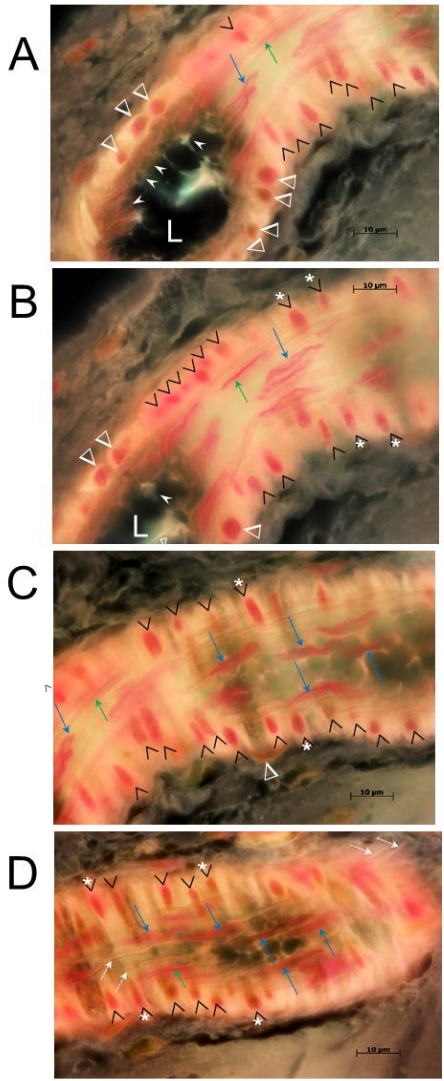

Figure 5 shows the optical sections of primo subvessel inside a primo node shown in Figure 4 C a. The diameter of the sinus is about 40 µm. The section plane of this slide passes through the lower part of the sinus along the length of the vessel and approximately 6 µm above the glass surface. The lower left portion of the first sinus exposes the lumen (L) (Figure 5 A) indicating that this part of the sinus is positioned in the section plane. Being in the section plane, the outer layer of the lumen shows nuclei of the fibroblasts of the longitudinal fiber (white empty arrowheads). These fibers go along the entire p-subvessels as shown in Figure 1 C (longitudinal fiber). The prominent nuclei of transverse fibers are visible further along the sinus length that look like parallel streaks and labeled with black arrowheads. These fibers are finer than the longitudinal fiber and have the appearance of coils around the sinus (Figure 1 C, transversal fiber). The nuclei of the muscle-like cells are very noticeable (blue arrow). These cells create the outer membrane of p-subvessel. It runs close to the longitudinal fibers. The nucleus of the endothelial cell is presented by the characteristic rod like structure (labeled by a green arrow). The layer of endothelial cells comprises the wall of p-subvessel. The layers of endothelial and muscle-like cells are shown in Figure 1, B. A few p-microcells (primo vascular microcells) attached to the internal surface of lumen are labeled by white arrowheads.

Figure 5. Optical sections of primo subvessel inside a primo node shown in Figure 4 C a. A. The optical section of p-subvessel is about one microns below the cutting plane of the vessel and perpendicular to the microscope objective axis. B. The optical section of p-subvessel is about one microns below the surface in A. C. The optical section of p-subvessel is about one microns below the surface in B. D. The optical section of p-subvessel is about one microns below the surface in C. White empty arrowheads– nuclei of the fibroblasts of the longitudinal fiber, white arrowheads - p-microcells, L – lumen, black arrowheads – nuclei of the fibroblasts of the fine transversal fiber, black arrowheads with asterisk – nuclei of the fibroblasts of the fine transversal fiber that belong to the same coil, blue arrow – nuclei of the muscle-like cells, green arrow – nuclei of endothelial cells, dotted arrows – thin ~1 µm fibers going along the length of the primo subvessel.

The optical section of the sinus that is positioned about one micron below the surface in panel A and slightly moved to the right along the sinus axis is shown in Figure 5 B. Being closer to the lower part of the external envelope, this section provides a better view of nuclei of the endothelial and muscle-like cells (green and blue arrows, respectively). The section also displays a few nuclei of the fibroblasts of the longitudinal fiber (white empty arrowheads) and many nuclei of the fibroblasts of the fine transversal fiber (black arrowheads). Some of the nuclei of the traversal fibers clearly belong to the same coils (black arrowheads with asterisk).

The next optical section is again one micron closer to the external envelope (Figure 5 C). The images of the nuclei of endothelial and muscle-like cells become more prominent than those in the previous optical section. Finally in the next optical section (Figure 5 D) that is very close to the sinus external envelope, the images the nuclei of endothelial and muscle-like cells become the dominant features of the section. They are parallel to the sinus axis and perpendicular to the pronounced nuclei of the traversal fibers. Thin ~1 µm fibers going along the length of the sinus were also noticed in this section (dotted arrows).

It is clear from the sections observed by high resolution light microscopy that fibers are very important for structural integrity of p-subvessels. The special attention to the extracellular fibers in primo-nodes and vessels was given in the investigation of matrices of primo-vascular system [22, 33, 53]. Summarizing and characterizing electron microscopy images of primo-vascular system from the previously published works, the authors confirmed the existence of two types of fibers [22]. These two groups are represented by thin collagen fibrils and thick non-collagenous wavy bent fibers. The diameters of the thick bent fibers were 30-200 nm, and they were highly bent. This group included organ surface primo-nodes and lymphatic primo-nodes and vessels. The diameters of the collagen fibrils were 30-70 nm. Organ surface, heart, and hypodermis primo-vessels were categorized in this group. Combining electron microscopy and variety of stains, Lee et al. [53] found that primo-subvessels surrounded by both the extracellular fibers and endothelial cells. It was suggested by Bonghan Kim [9] and confirmed by K.S. Soh research group [36] that the rod-shaped nuclei aligned along the major axis of p-subvessels have been accepted as one of the characteristics of those endothelial cells. Though, Lee et al. [53] suggested that those rod-shaped nuclei may belong to the extracellular fibers. Nevertheless, this issue needs further investigations because apparently the same rod-shaped nuclei were recently found in endothelial cells of the primo-vascular system of the rat spinal cord [33].

The general structural architecture of the p-subvessels and sinuses observed by high resolution microscopy agrees well with those described by Kim [9]. It is also consistent with results obtained by electron [25, 36, 53] and light [32, 34, 35] microscopies.

## P-microcells are progenitors of multipotent stem cells

Kim revealed in his pioneer work that p-microcells carry DNA, could proliferate in culture, can circulate in PVS system, regenerate injured tissue, and overall behave as multipotent stem cells. [9].

Figure 6 shows the optical sections of a broaden p-subvessel (sinus) inside a primo node shown in Figure 4 C b. Panel A depicts the image of the sinus in bright field mode in order to see proper colors produced by the hematoxylin eosin staining. The sinus envelope displays a nucleus of the fibroblasts of the longitudinal fiber (black empty arrowheads), nuclei of the fibroblasts of the transversal fiber (black arrowhead), and nuclei of endothelial cells (green arrows). They all are stained red by eosin [9, 57-59].

The sinus is filled with p-microcells of 1-4 µm. Many of microcells are of irregular shape (black arrow), but a few of the cells are round as shown by arrows with oval heads. Many cells are grouped and connected by thin processes (exemplified by processes labeled a black filled arrowhead). Hematoxylin colors p-microcells with blue-purple hue. The basophilic structures containing nucleic acids, such as the ribosomes and the chromatin-rich cell nucleus, and the cytoplasmic regions rich in RNA are known to be stained with blue-purple hue by hematoxylin [59]. These results agree well with those obtained with primo nodes removed from the surfaces of rat internal organs. The samples were fixed with formalin, embedded in paraplast, sliced and stained with hematoxylin-eosin [16]. Authors indicated the presence of "small stem-like" cells of 3-4 µm in diameter, that express stem cell marker CD133 (antigen encoded by the PROM1 gene) and pluripotent transcription factors such as Oct4 (octamer-binding transcription factor 4 encoded by the POU5F1 gene) and Nanog (transcription factor encoded by the NANOG gene).

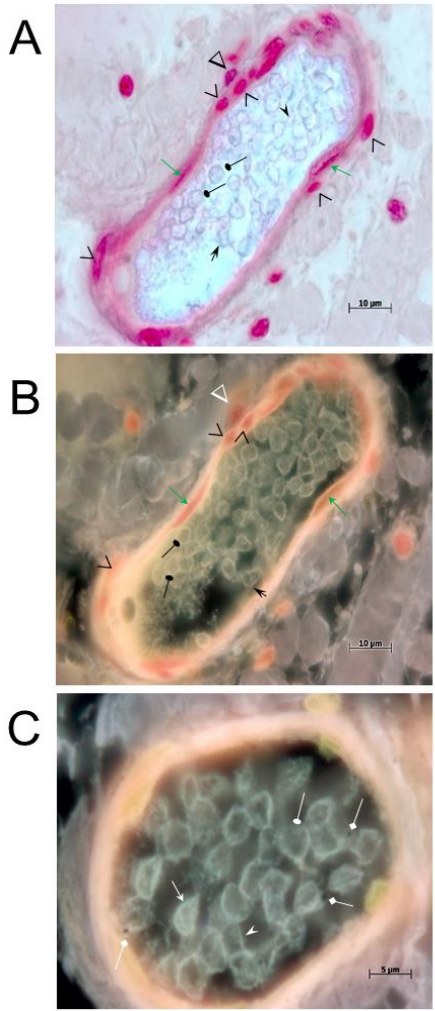

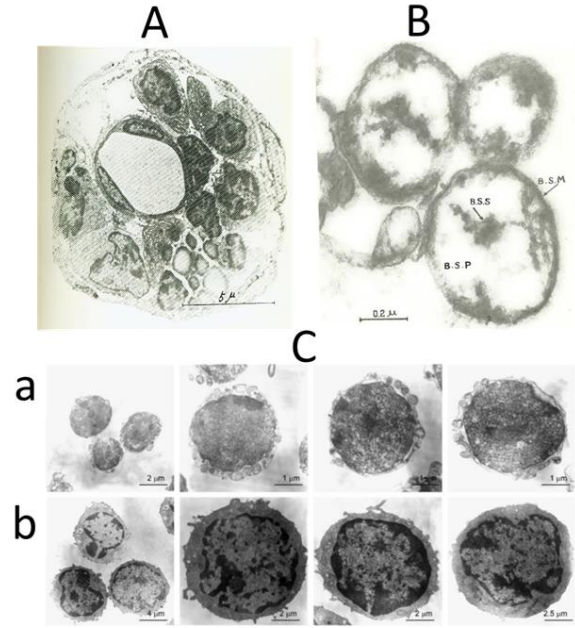

Figure 6. Optical sections of p-subvessel inside a primo node shown in Figure 4 C b. A. The bright field image of the optical slide positioned about 1 µm below the surface and perpendicular to the microscope objective axis. Black arrow - cells of irregular shape, arrows with oval heads - round cells, black filled arrowhead - thin processes, black empty arrowhead– nucleus of the fibroblasts of the longitudinal fiber, black empty arrowheads – nuclei of the fibroblasts of the fine transversal fiber, green arrow – nuclei of endothelial cells. B. The darkfield image of the same optical section shown in A. The labels are the same as in panel A. The magnified image of p-subvessel sinus taken from the 69th slide of the same primo node as shown in Figure 10. The optical sections imaged at about 2 µm below the wall cross-section. The p-microcells are of irregular (arrow) and round (oval head arrow) shapes. Some cells are linked with processes (arrow head). Cell sizes vary from 1 to 4 µm. A few round cells with black center are labeled by square head arrows.

Figure 7. TEM micrographs of p-microcells, very small embryonic-like (VSEL), and hematopoietic stem cells. A. External primo node. Star – p-microcells. B. P-microcells. BSS - p-microcell nucleosome, BSP - p-microcell nucleoplasm, BSM -= p-microcell membrane [9]. C. TEM of very small embryonic-like (VSEL) cells and hematopoietic stem cells. (a) small embryonic-like (VSEL) cells are small and measure 2-4 µm in diameter. They possess a relatively large nucleus surrounded by a narrow rim of cytoplasm. The narrow rim of cytoplasm possesses a few mitochondria, scattered ribosomes, small profiles of endoplasmic reticulum and a few vesicles. The nucleus is contained within a nuclear envelope with nuclear pores. Chromatin is loosely packed and consists of euchromatin. (b) In contrast hematopoietic stem cells display heterogeneous morphology and are larger. They measure on average 8–10 µm in diameter, possess scattered chromatin and prominent nucleoli (Reprinted by permission from Macmillan Publishers Ltd: Leukemia[60], copyright, 3565580395398, 2006).

The darkfield image of the same optical section shown in a panel B of Figure 6. The image likewise presents p-microcells of irregular and round p-microcells inside the sinus. The nucleus of the fibroblasts of the longitudinal fiber, nuclei of the fibroblasts of the fine transversal fiber, and nuclei of endothelial cells are also shown. The magnified image of p-subvessel sinus taken from a different slide of the same primo node is shown in Figure 6 C. The optical section is focused at about 2 µm below the wall cutting plane. The p-microcells are clearly visible in this image. The p-microcells observed by high resolution light microscopy compare well with p-microcells obtained by Kim and with small embryonic cells (VSEL) shown in Figure 7 [13].

Figure 7 A shows p-microcells within a small external primo node. Kim carefully detailed p-microcells found within the node. In the cytoplasm of the cells, one nucleus or two nuclei are located in the center or periphery of the cell. These nuclei have thin membranes and plentiful chromatin, and the nucleoli can be found lower in the center of the nuclei. The nuclear membrane sometimes is wrinkly. Bordering cells appear near to each other or are connected to each other through small cytoplasmic processes. Isolated p-microcells at higher magnification are revealed in Figure 7 B. Kim observed that the p-microcell was typically spherical but frequently oblong in shape. An average p-microcell was 1.2-1.5 microns in size, while the smallest one appearing 0.8 micron and the biggest one, which was rarely viewed, was 2.4 microns. P-microcell had a thin membrane of high density with very distinctive contour. The nucleus took various forms. The sanal membrane was very dense and had a distinctive shape. Granules of various sizes and relatively high electron density were seen in the cytoplasm close to the p-microcell membrane. Figure 7 C shows a TEM of very small embryonic-like (VSEL) cells and hematopoietic stem cells extracted from murine bone marrow by multi-parameter sorting [60]. We noticed (Figure 7 C) that very small embryonic-like (VSEL) cells, as compared to hematopoietic stem cells, were smaller in size (2-4 vs. 8-10 µm), have comparatively large nuclei and a thin rim of cytoplasm. The authors hypothesized that this population of very small embryonic-like (VSEL) stem cells was deposited early during development in bone marrow and could be a source of pluripotent stem cells for tissue/organ regeneration[60]. It should become noticed that p-microcell are very similar by morphology to these cells [13].

The earliest electron microscopy data were obtained with p-vessels and p-microcells from peritoneum of rabbit and human umbilical cord.

The TEM images of p-microcells revealed the p-microcells to be of irregular shape and 1-2.5 µm in size [19]. A more detailed morphological analysis of p-microcells was carried out with p-microcells obtained from p-vessels and nodes harvested from surfaces of the rabbit internal organs [26]. The authors found that p-microcells were 1.7-2.5 µm in diameter. They had a small nucleus surrounded by a layer of cytoplasm and a plasma membrane. TEM also revealed that they had "cytoplasmic protrusions like pseudopodia." They also have fragmented DNA [26]. The authors stated that p-microcells were different in morphology from apoptotic bodies and bacteria. The "cytoplasmic protrusions" observed in this work are very similar to the "processes" described by Kim (Figure 7, A and B) and observed in our work (Figure 6). Further morphological analysis of p-microcells were carried out by atomic force and electron microscopies [18, 20, 23]. EM images of p-microcell obtained from the intra- lymphatic p-nodes and vessel of mice [18] have morphological features similar to those obtained by Kim and shown in Figure 7 [13].

The stem cell nature of p-microcells was fully described by Kim [6, 9]. In 2005, Soh and his group put forward the hypotheses that resonated with Kim's ideas on the stem cell character of p-microcells. They suggested that the DNA granules that move in the PVS are naturally produced p-microcells. They can grow and differentiate into a variety of tissue cells, as well as regenerating the tissues of injured organs [61]. That is to say that p-microcells are progenitors of multipotent stem cells. It was a very fruitful idea. It reinstated Kim's discoveries and laid out a course for future characterization of the p-microcells with the use of stem cells markers [15-18, 51, 62]. The sliced primo vascular vessels and nodes obtained from the internal organ surfaces of rat were immunostained by Integrin beta 1 (binding protein 1 encoded by the ITGB1BP1 gene), Collagen type 1 (expressed by COL1A1, COL1A2 genes), Fibronectin (high-molecular weight glycoprotein of the extracellular matrix), CD54 (protein encoded by the ICAM1 gene), Thy 1, (glycophosphatidylinositol) and vWF (Von Willebrand factor) antibodies. Both hematopoietic and mesenchymal stem cell markers were found to be strongly expressed similarly to that of the stem cells in bone marrow [15]. Microcells extracted from the primo vascular nodes obtained from the surface of rat liver, were analyzed by RT-PCR (reverse transcription polymerase chain reaction) [17] and reported to express the pluripotent stem cell markers, Oct4, Sox2, Stella, Rex1, and Klf4. Similarly, primo node microcells removed from the inside of the mice blood and lymphatic vessels, appear to be 3-5 µm in diameter and express Oct4, Nanog, SSEA-1, and Sox2 stem cell markers [18].

## Conclusions

1. The primo-vascular system is an essential component of the circulatory system, together with blood vessels and lymph vessels. The vascular cell system is distributed throughout the entire body, over and inside of various organs, inside and outside of the blood and lymphatic vessels, in the internal and the peripheral nervous system, and in the corium or in the subcutaneous layers of the skin. Two structural elements - vessels and nodes comprise this system.
2. The primo nodes are various shapes elements (round, oval, or multifaceted). A vessel bundle of the incoming (afferent) vessels enter into the node, branch into additional bundles, and fill the node interior by tightly spun and folded bundles. Subvessels converge and exit from the node as efferent primo vessels. The enlarged p-subvessels inside the node, which are called the sinuses of the node, harbor microcells, the progenitors of multipotent stem cells.

## Acknowledgments

We thank Mary Rudisill, and David Pascoe for critical reading of the manuscript. Supported by Lake Erie College of Osteopathic Medicine, School of Kinesiology and College of Veterinary Medicine, Auburn University.